\documentclass[utf8]{FrontiersinHarvard} %
\usepackage{url,hyperref,lineno}
\usepackage[onehalfspacing]{setspace}
\usepackage{xcolor}
\usepackage{float}

\def\keyFont{\fontsize{8}{11}\helveticabold }
\def\firstAuthorLast{} 
\def\Authors{Maisha Islam Sejunti  
    \,$^{1,*}$, Melissa Butler\,$^{1}$, Yingjie Hu\,$^{2}$ and Dane Taylor\,$^{1}$}
\def\Address{$^{1}$School of Computing and Department of Mathematics \& Statistics, University of Wyoming, Laramie, WY, 82072, USA. \\
$^{2}$Department of Geography, State University of New York at Buffalo, Buffalo, NY, 14260, USA.  }
\def\corrAuthor{Maisha Islam Sejunti}
\def\corrEmail{msejunti@uwyo.edu}

\begin{document}

\onecolumn
\firstpage{1}

\title[Predictability of Human Movements]{Predictability of Human Movements across Industry Sectors using Multilayer  Networks} 

\author[\firstAuthorLast ]{\Authors} 
\address{} 
\correspondance{} 
\extraAuth{}

\maketitle
\begin{abstract}
Predicting the patterns of human movement is important to applications from urban design to disease control and emergency response. In this work, we leverage machine learning (ML) to study the predictability of movements to different types of locations (e.g., movements to grocery stores, to schools, or to hospitals). Our approach relies on developing multilayer mobility networks encoding movements between  census tracts with network layers associated with different industry sectors. We study the predictability of node degrees across layers, studying which industries are easier/harder to predict. Rather than treat prediction as an end in itself, we  propose ML predictability as an exploratory tool to understand multilayer network organization. Our results suggest that nodes' out-degrees are easy to predict, since they are driven by census tract populations; whereas in-degrees are harder to predict, since persons can be drawn to diverse type of locations in a census tract. We also find that movements to certain industry types (e.g., restaurants) are easier to predict than others, and the different industries' predictability are positively  correlated with the number of movements to those types of locations. These findings provide a practical step toward  using ML for multilayer mobility modeling and the many downstream application.
\tiny
 \keyFont{ \section{Keywords:} Human mobility, Multilayer networks, Industry sectors, Machine learning models, Predictability} 
\end{abstract}

\section{Introduction}\label{sec:intro} 

Human mobility research involves studying how  people interact with their spatial environment \citep{gonzalez2008understanding,louail2015uncovering,pappalardo2023future}. An improved understanding for movement patterns  supports many downstream  practical applications including urban planning \citep{gao2015spatio,jahromi2016simulating,yang2016exploring}, 
transportation system design \citep{li2025understanding,kitamura2000micro}, 
mitigating contagious diseases \citep{meloni2011modeling,tizzoni2014use,belik2011natural,yang2011impact}, 
emergency response 
\citep{huang2026analysis,wang2014quantifying,han2019cities}, 
and  economic growth  planning \citep{fujita2001spatial,schlapfer2021universal}. 

Notably, recent work has focused on  integrating mobility data  into  machine learning (ML)  models   to predict movements  
\citep{cuttone2018understanding,kim2018method,toch2019analyzing,wang2019urban,luca2021survey}, supporting urban planning \citep{lim2026spatiotemporal}, predicting income levels \citep{gao2024income}, supporting public health  \citep{rustamov6315262rural,wang2025analyzing,rahman2021machine}, and enhancing crime prevention
\citep{wu2022enhancing}. Nevertheless, predicting node-level structural properties, such as in-degree and out-degree, remains an underexplored topic for multilayer mobility networks---particularly those with layers encoding movements to locations associated with different industries.

Here, we develop predictive models for human movements that are stratified into industry-related categories: movements to schools, to restaurants, to hospitals, and so on. Our approach builds on our prior work \citep{butler2026multilayer} that utilizes multilayer networks where nodes represent spatial regions (e.g., census tracts or census block groups) and edges in  different layers  represent observed movements associated with different industry sectors. We note that the layers of multilayer mobility networks have also been used to represent different modes of transportation \citep{chodrow2016demand, de2014navigability, parshani2010inter, halu2014emergence, taylor2015topological} (e.g., roadways, metro lines, airlines, etc.); however, we focus on using them to stratify movements based on the category for the movement destination. Of note, we also consider weekly movement changes so that the industry-stratified movement data is represented by time-varying, multilayer, mobility networks.

\begin{figure*}[t]
    \centering
\includegraphics[width=\linewidth]{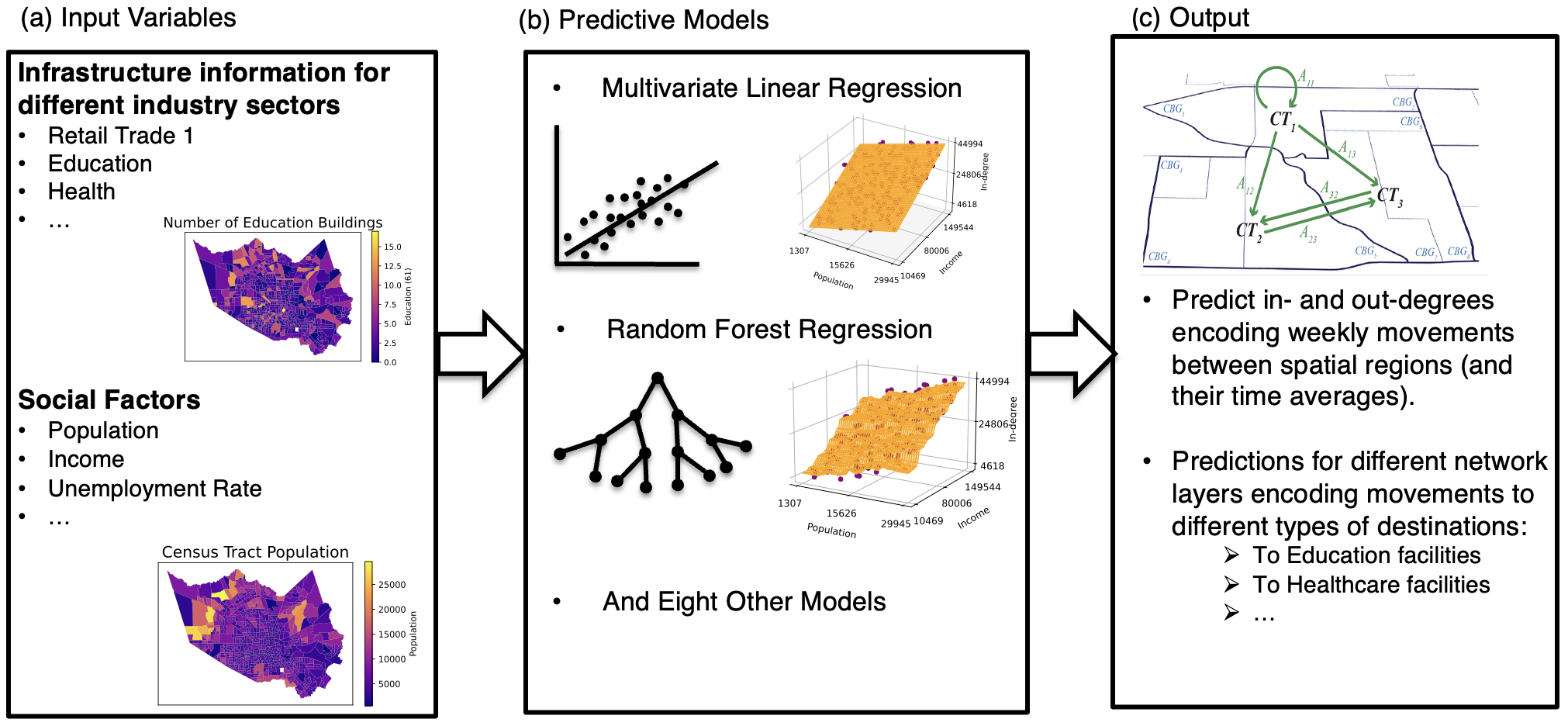}
    \caption { \textbf{Machine-learning (ML) framework studying predictability of human movements to locations associated with different industries.} 
    (a) Input data includes both infrastructure information derived from the Safegraph dataset \cite{safegraphdata}---that is, the number of buildings in each spatial region for each industry types---as well as demographic and Socioeconomic information obtained from the U.S. Census. For additional discussion on the infrastructure data and demographic/Socioeconomic data, see Secs.~\ref{sec:multilayer}  and Sec.~\ref{sec:two_models}, respectively.
    (b) We develop and compare ten statistical and machine learning (ML) models including
    multivariate linear regression (MLR) and random forest regression (RFR) to quantify how accurately mobility can be predicted across industry types. See Secs.~\ref{sec:two_models} and  \ref{sec:industry} for comparison of these two models and Secs.~\ref{sec:Survey} and \ref{sec:Survey_Target} for comparison across all ten models. 
    (c) We predict weekly and time-averaged movements in and out of census tracts as encoded by nodes' in- and out-degrees for mobility networks. We also stratify the movements according to their destinations' industry type (e.g., movements to schools,  hospitals, etc.), allowing us to identify which types of movement are more or less predictable. See Sec.~\ref{sec:multilayer} for discussion on these time-varying, multilayer mobility networks.
    } 
    \label{Schematic_Figure}
\end{figure*}

Rather than treating prediction as an end in itself, we use predictive modeling as a scientific tool to investigate the structural predictability of these empirical mobility networks. Specifically, we ask which aspects of node connectivity can be predicted from measurable characteristics of the underlying spatial regions, and whether the degree of predictability varies systematically across industry-specific network layers, movement directions, and temporal scales.  

To quantify network predictability, we use established statistical and ML models to relate node-level connectivity to demographic, socioeconomic, infrastructure, and temporal characteristics of the corresponding spatial regions. 
We predict nodes’ in- and out-degrees, which encode observed movements in and out of spatial regions.
In Figure~\ref{Schematic_Figure}, we present an overview of our methodology that includes descriptions for the (a) input variables, (b) predictive models, and (c) outputs for different predictive tasks. 
As illustrated in Fig.~\ref{Schematic_Figure}(a), we train predictive models based on two types of input data: infrastructure information about the number of buildings of particular industry types in each spatial region as well as demographic and Socioeconomic information. As illustrated in Fig.~\ref{Schematic_Figure}(b), we  compare ten predictive models including multivariate linear regression and random forest regression.
As illustrated in Fig.~\ref{Schematic_Figure}(c), we study several node-degree-related prediction tasks including weekly  movements into and out of census tracts   as well as their time-averaged values. Prediction tasks also include making estimates for both industry-specific movements in/out of regions as well as the total movements in/out of regions (i.e., as encoded by the layer-aggregated network associated with the summation of the layers' adjacency matrices). Comparing these prediction tasks allows us to examine how structural predictability changes across industry layers, movement directions, and temporal scales.

These experiments give rise to several insights. 
%
%
First, we find that   in-degree prediction is generally more challenging than   out-degree prediction, which is true across industry types and for both weekly and time-averaged node-degree predictions. We find that a driving factor is that outward movements are less complex, being driven by census tracts' population sizes, whereas inward movements are more more complex, since persons are drawn to diverse types of locations.
We  identify the most influential predictors (i.e.,  population for out-degree prediction and the amount of sector-specific infrastructure for in-degree prediction) to be consistent when considering either the regression coefficients for multivariate linear regression or importance measures for random forest regression.

By studying the predictability of movements to destinations having different industry types, we also identify which types of movements are easiest to predict (that is, movements to locations associated with food and retail) and hardest to predict (movements to locations associated with education and other services). One driver for the differing predictability across industry types appears to be the total the number of observed movements  each  industry type, with more-accurate in-degree predictions for industry locations that attract more movements. Our comparison of ten predictive models for weekly node-degrees also reveals that the nonlinear predictive models outperform the linear ones; they obtain up to a 36\% increase in prediction accuracy for out-degree prediction (15\% for in-degree prediction) when time-encoding variables are included as an input variables.

Overall, we develop node-degree prediction as a ML tool to study the organization of multilayer mobility networks---particularly those in which different layers represent movements to locations associated with different types of industries. Our findings expand our understanding of structure prediction for multilayer networks, and they provide practical insights that can support downstream applications that may include urban planning and disaster response.

This paper is organized as follows.
In Sec.~\ref{sec:data}, we discuss our data and methodology.
In Sec.~\ref{sec:results}, we present our main results.
A discussion and conclusions are provided in Sec.~\ref{sec:discuss}.

\section{Data and Methodology}\label{sec:data} 

Here we discuss the datasets supporting our study and the methodology that we use to predict industry-stratified movements between spatial regions.
In Sec.~\ref{sec:multilayer}, we discuss the Safegraph dataset that provides both human mobility data and infrastructure information across industrytypes.
In Sec.~\ref{sec:social}, we discuss demographic and socioeconomic data derived form the U.S. Census.
In Sec.~\ref{sec:predict}, we present the statistical and machine learning methods that we use to make predictions.
In Sec.~\ref{sec:entropy}, we discuss the entropy analysis method for the time-series complexity.

\begin{figure}[b]
    \centering
    \includegraphics[width=\linewidth]{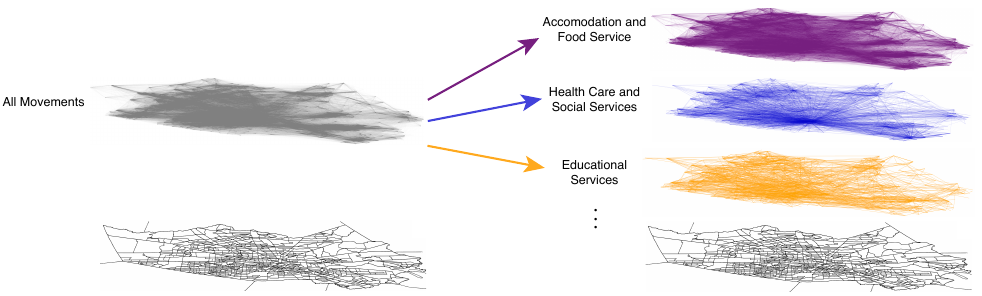}
    \caption{\textbf{Multilayer mobility networks stratified by industry type.} Illustration of mobility networks, stratified by industry types, using the North American Industry Classification System (NAICS). The top network depicts all movements, whereas the lower networks depict movements to locations associated with a particular industry type: accommodation and food service (NAICS code 71), health care and social services (62), and educational services (61). Our study focuses on Harris County, Texas, and a map is shown illustrate the boundaries of census tracts.}
    \label{fig:maps}
\end{figure}

\subsection{Infrastructure Data and Industry-Stratified Multilayer Mobility Networks} \label{sec:multilayer}


Our study is based on the the SafeGraph dataset \citep{safegraphdata} (approximately 700 GB), which contains weekly human movement patterns derived from cell phone GPS locations. The dataset records weekly movements from individuals' home locations, each represented by a home census block group, to destination locations. Each destination location (commonly called a ``point of interest'') is associated with a unique location identifier that integrates a geospatial encoding system and provides a set of attributes including the location name, latitude and longitude, an industry classification, and several business details (if available). In particular, each destination location is categorized according to the 
2017 North American Industry Classification System (NAICS), allowing us to group the locations into their industry types such as  education services (NAICS code 61), health care (62),  food services and accommodation (72), and so on. Finally, we note that the SafeGraph dataset is no longer maintained for real-time analysis, and a successor dataset called Advan Research Foot Traffic can be purchased from Dewy Data.

Following the methodology developed in \citep{butler2026multilayer},
we construct and study time-varying, multilayer mobility networks  in which nodes represent census tracts---each of which is comprised of several census block groups---and each weighted edge represents the weekly count of observed movements from one census tract to another such that the persons' destinations are of a particular industry sector. That is, we let different network layers encode movements according to different industry types as illustrated in Fig.~\ref{fig:maps}.
%
More specifically, we build  time-varying adjacency matrices, $\textbf{A}^{(n)}(t)\in\mathbb{R}^{C \times C}$, where $t\in\{1,\dots, T\}$ is an index identifying a particular week of the study duration, $n$ is an index identifying a particular NAICS code, and $C$ is the total number of census tracts under consideration. We enumerate the census tracts $i\in\{1,\dots,C\}$.  It follows that $\textbf{A}_{ij}^{(n)}(t)$ denotes the number of observed movements from the i-th to the j-th census tract during week $t$ for which the destination has industry type $n$. (We note that such adjacency matrices have also been called `origin-destination' matrices.)

Note that if we define $\textbf{B}_{pq}^{(n)}(t)$ to be the number of observed movements during week $t$ from census block group $p$ to a destination $q$, which has an associated NAICS code $n$, and  we further define $H_i$ as the set of CBGs within the i-th census tract $i$ and $P_j^{n}$ as the set of destination locations within the j-th census tract with NAICS code $n$, then the adjacency matrix is given by
$\textbf{A}^{(n)}_{ij}(t)=\sum_{p\in H_i, q\in P_j^{n}} \textbf{B}_{pq}^{(n)}(t).$
It is also worth noting that $|P_j^{n}|$ (i.e., the number of elements in set $P_j^{n}$) gives the number of locations in j-th census tract associated with NAICS code $n$. We will later use this information to help train machine learning models to predict movements in and out of census tracts. We also note that this notation allows us to define the total number of observed movement in each network layer $n$ during week $t$  as $M^{(n)}(t)=\sum_{ij}\textbf{A}^{(n)}_{ij}(t)$, and we let  $\overline{M^{(n)}}=\sum_{t}M^{(n)}(t) $ denote the time average.

Throughout this paper, we will also study the associated layer-aggregated network that encodes all movements types and has a 
time-varying adjacency matrix
$\textbf{A}(t)=\sum_n \textbf{A}^{(n)}(t).$
We note in passing that the opposite mathematical operation---that is, considering an initial network that encodes all movement types and then separating it into different network layers---is commonly referred to as network ``stratification''. Therefore, we refer to this type of multilayer network as an ``industry-stratified'' multilayer mobility network.

To further align with and compare to prior insights made in  \citep{butler2026multilayer}, we will similarly focus on predicting movements within Harris County, Texas. That said, we will consider a longer time period from  Monday, January 8, 2018 to Monday, December 30, 2019 that contains 103 weeks.
In contrast, Ref.~\citep{butler2026multilayer} studied the impact of a deadly winter storm and focused only on the storm week and the six preceding weeks. We also note that our work contrasts \citep{butler2026multilayer}, since that paper did not consider machine learning models nor did it consider movement predictability across different industry types.

To study human movements across industry types, we will focus on the coarsest level of NAICS categories, and we will focus our attention on the eight NAICS codes that were found to be associated with the most movements: retail trade 1 (44), retail trade 2 (45), real estate (53), education (61), health care (62), arts, entertainment (71), food services and accommodation (72), and other services (81). The 2-digit numbers after the categories represent their corresponding NAICS codes. 

\begin{figure}[t]
    \centering
    \includegraphics[width=1\linewidth]{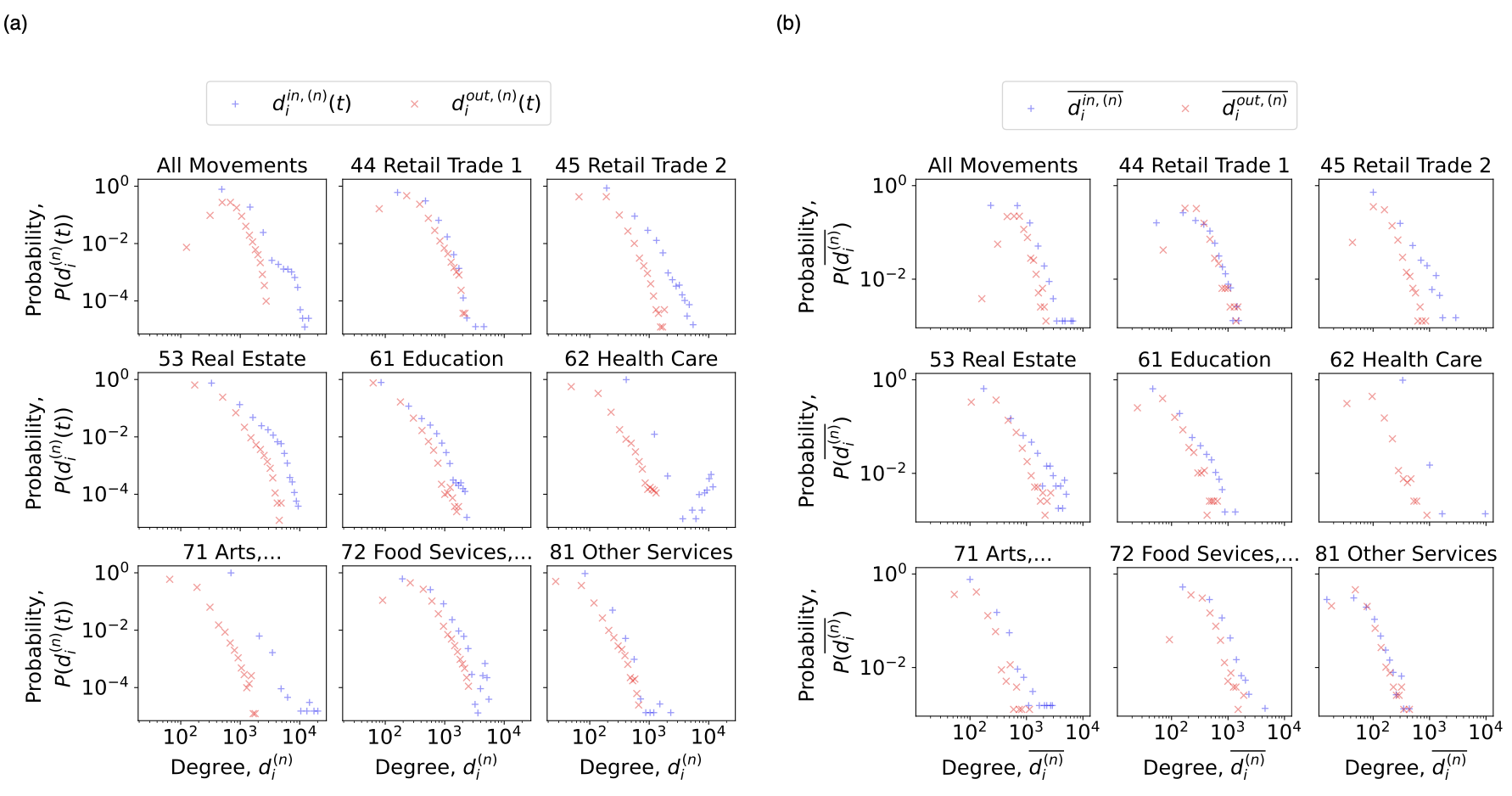}
    \caption{
    \textbf{Distributions of weekly and time-averaged node degrees across census tracts for different industry types.} 
    Blue and red markers represent in- and out-degrees, respectively. The distributions are heavy tailed across all network layers, and the in-degrees are typically more heterogeneous.
    }
    \label{observed_mlr_mean}
\end{figure}



The primary aim of this present paper is to  predict nodes' in- and out-degrees across the multilayer network. In the context of human mobility, a census tract's in-degree quantifies inbound movement, whereas its out-degree quantifies outbound movement (each stratified by the destination's industry type).  Specifically, for each network layer $n$ and each week $t$, we let  
    $d_i^{in,(n)}(t)=\sum_j \textbf{A}_{ji}^{(n)}(t)$ and 
    $d_i^{out,(n)}(t)=\sum_j \textbf{A}_{ij}^{(n)}(t)$
denote the in- and out-degrees for the i-th census tract (sometimes called the  in/out ``intralayer'' node degrees, since we are summing edges' weights within a layer). Similarly, we let 
    $d_i^{in}(t)=\sum_n d_i^{in,(n)}(t)$ and
    $d_i^{out}(t)=\sum_n d_i^{out,(n)}(t)$
denote the in- and out-degrees for the layer-aggregated network. Note that each node degree can change weekly, and so we also define their time-averaged values by $\overline{d_i^{in,(n)}}$, $\overline{d_i^{out,(n)}}$, $\overline{d_i^{in}}$ and $\overline{d_i^{out}}$.


In Fig.~\ref{observed_mlr_mean}, we depict distributions of nodes' weekly and time-averaged in-degrees (blue $+$ symbols) and out-degrees (red $x$ symbols) across the $786$ census tracts in Harris County, Texas, for the layer-aggregated network and the eight network layers associated with different industry types. Observe all distributions are heavy-tailed, 
highlighting the significant heterogeneity in mobility patterns across the census tracts; most have a small amount if movement in and out, whereas it is very large for a few census tracts.
This heterogeneity is typically larger for the in-degrees. 
For example,  for the network layer associated with healthcare, we should expect   $d_i^{in,(n)}(t)$ and $\overline{d_i^{\text{in},(n)}}$ for the  census tracts  containing hospitals, but they should be are very small for the other census tracts. 

\subsection{Demographic and Socioeconomic Data} \label{sec:social}
To develop predictive models for movements in and out of census tracts, we will utilize two types of input features: the counts of industry-specific infrastructure in each census tract (recall the variable $|P_j^{n}|$ defined above) as well as demographic and socioeconomic information for them.
%
These are derived 
from the U.S. Census Bureau’s  American Community Survey (data in 2019), and we focused on
six social factors: population (B01003), population density, income (B19013), unemployment rate (DP03), percentage of non-white (B02001), and poverty rate (S0601). (Here the number corresponds to a U.S. Census feature identifier.) Notably, for Harris County, TX, these were previously shown to contain complementary, non-redundant information  using a set of variance inflation factor (VIF) tests that were applied to a larger set of features \citep{butler2026multilayer}. (See Fig.~\ref{Schematic_Figure} for visualizations  of how education infrastructure and population  vary across the census tracts in Harris County, TX.)
%



\subsection{Predictive Models for Node Degrees}\label{sec:predict}

Our main contribution is the development, analysis and comparison of predictive models for in- and out-degrees for network layers of the industry-stratified multilayer mobility network (as well as its associated layer-aggregated network).
%
Our approach relies on training statistical and machine learning models that integrate both the industry infrastructure information across our selected NAICS codes as well as demographic and socioeconomic data from the U.S. Census.

We will compare ten statistical and machine learning models. Initially, we will compare multivariate linear regression and random forest regression, studying their predictive accuracy and ranking the importance of different input features for these models.
%
Letting $\{y_{i}\}_{i=1}^{n}$ denote a set of node degrees to predict and $\{X_{ij}\}$ denote its associated set of predictor variables enumerated by $j\in \{1,\dots,K\}$, multivariate linear regression (MLR) takes the form \citep{helwig2017multivariate}
$y_{i} =b_{0}+\sum_{j=1}^K b_{j}X_{ij} + \epsilon_{i}.$
Here, $\epsilon_i$ represents the model error and the regression coefficients
$b_{j}$ are fit to minimize the mean square error, $\sum_i \epsilon_i^2$.
%
In our case,  for $j\in\{1,\dots, 8\}$ the $X_{ij}$ variables represent infrastructure information across the eight NAICS categories that we study. For $j\in\{9,\dots, 14\}$, the variables represent the above-mentioned social factors from the U.S. census.

Across all prediction tasks, model accuracy is measured by R-squared values, also known as the coefficient of determination for , which measures how well the model explains the variability of the dependent variable.
\begin{equation}
R^2=1-\frac{\text{Residual sum of squares}}{\text{Total sum of squares}}=1-\frac{\sum_i(y_i-\hat{y}_i)^2}{\sum_i(y_i-\bar{y})^2}.
    \label{r_squared_value}
\end{equation}
Here, $y_i$ is the observed value of the dependent variable at the $i-th$ data point and $\hat{y}_i$ is the predicted value for the corresponding observed value, and $\bar{y}$ is the average of all the values $y_i$. If $R^2=1$, the model explains all variance and $R^2=0$ indicates that the model does not explain any of the variance.


We also evaluated model performance using root mean square error (RMSE), mean absolute error (MAE), and mean absolute percentage error (MAPE). RMSE and MAE are expressed in the same units as the target variable and therefore depend on the scale of the data. Since the mean in- and out-degrees differ substantially across the census tracts and across NAICS categories (recall Fig.~\ref{observed_mlr_mean}), RMSE and MAE are not well suited to make comparison among industrytypes. We also found MAPE to be inferior to R-squared values; many of the node degrees are zero or very small, causing the percentage errors to be unbounded and sometimes infinite. 
In contrast, $R^2\in(0,1)$ is stable and scale independent, and we found that it provided the most interpretable and intuitive insights across experiments. See \cite{chicco2021coefficient} for a broader discussion.





The second model we consider is  random forest regression (RFR), which
is a collection of decision trees that are constructed from different subsets of the input data and combine to provide an overall regression estimate \citep{cutler2007random}.
We implement RFR using the Scikit-Learn Python package and identify the method's optimal hyperperameters by performing a grid search (implemented using GridSearchCV in the Scikit-learn Python package) to identify the best number of estimators, the maximum number of features (which are randomly selected at each split), the maximum depth for the decision trees, and the minimum number of samples for each leaf in the decision tree \citep{ali2023case}. The parameter choices that we use to conduct the grid search are indicated in Table~\ref{ML_table}.

\begin{table*}[t]
\caption{\textbf{Predictive models and their searched hyperparameters.} Each   is categorized as linear/non-linear. The set following each hyperparameter indicates the values considered during the grid search to identify optimal parameters. }
\label{ML_table}
    \centering
\includegraphics[width=1\linewidth]{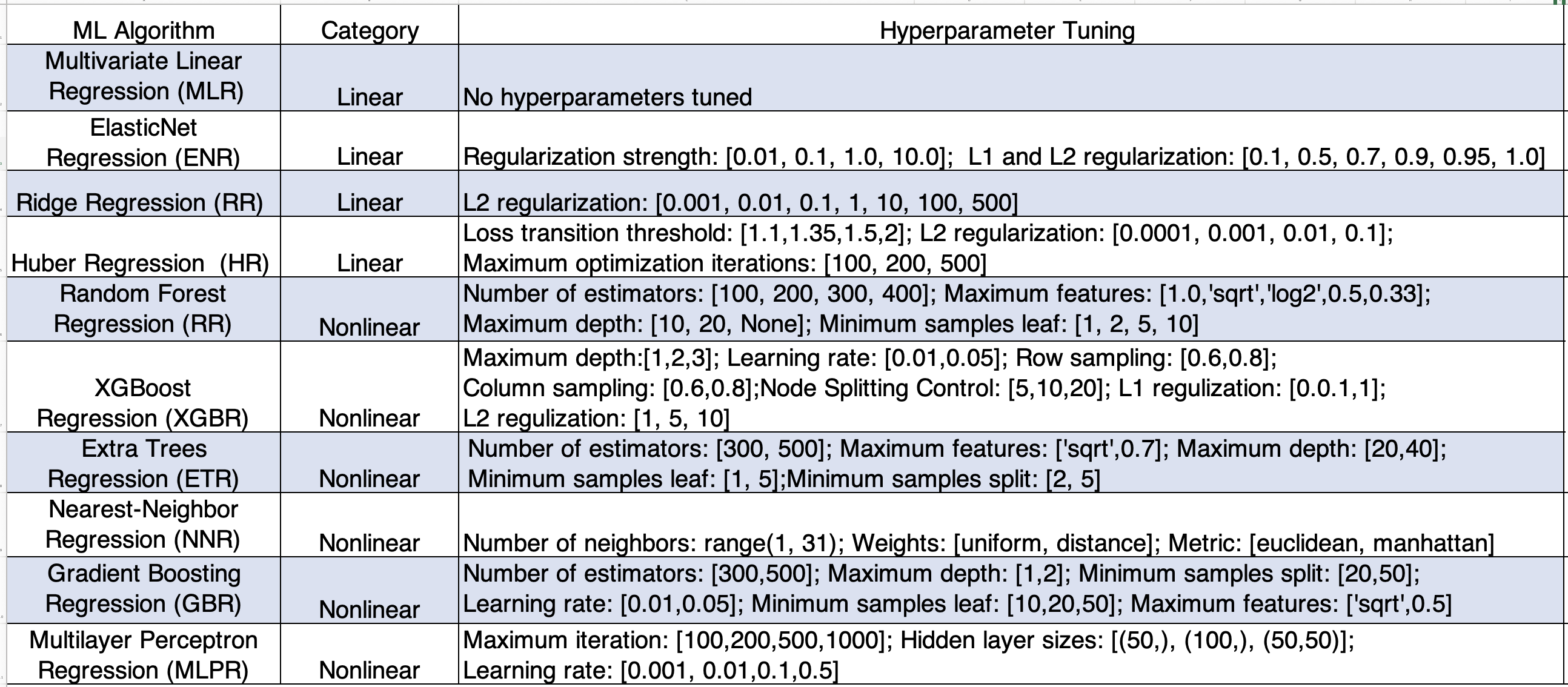}
\end{table*}

To train the models, we partitioned the dataset into training and testing sets using an 80:20 split, implementing
a five-fold cross-validation scheme.
During each iteration, 80\% of the data (i.e., four of the folds) was used for model training and 20\% (i.e., one fold) was used as a testing set. This process was repeated five times, so each fold operated as the testing set during one of the five trials. The average performance across the five testing folds was utilized to select the best hyperparameters during the grid search procedure. 

For both the MLR and RFR models, we study the importance of the different input features that enable prediction. For MLR, features' importance is quantified by their associated regression coefficients, and the largest-in-magnitude coefficients are interpreted as the most important ones. (This interpretation requires that the input data are normalized.) For RFR, we quantify feature importance through studying the impurity decreases across the decision trees. 
%
That is, each feature importance is determined by assessing how much each feature, on average across all trees in the forest, reduces impurity in the nodes where it is used. This reduction is computed as a weighted average, with each node weighted by the number of training samples it contains. During training, Scikit-Learn automatically calculates these importance scores for all features and normalizes them so that their total sums to one.



In addition to MLR and RFR, we study eight other machine learning models including three other linear models---elasticNet regression (ENR), ridge regression (RR), Huber regression (HR)---and five other nonlinear models---XGBoost regression (XGBR), extra trees regression (ETR), nearest-neighbor regression (NNR), gradient boosting regression (GBR), and multilayer perceptron regression (MLPR).
All models were implemented using Sci-Kit Learn and grid searchers were implemented using GridSearchCV. In Table.~\ref{ML_table}, we provide a list of these models and their corresponding hyperparameters and their grid search ranges. When training each model, we applied the same 80:20 split for training and testing data as well the 5-fold cross-validation procedure.

\subsection{Entropy Analysis for Time-series Complexity}\label{sec:entropy}

For each network layer $n$, we consider the set of time series $\{M^{(n)}(t)\}$ for $t\in\{1,...,103\}$, where $M^{(n)}(t)$ represents the total number of observed movement in layer $n$ at week $t$. To quantify the complexity of movement patterns across different industry sectors, we estimate the probability distribution of $M^{(n)}(t)$ using a histogram with $50$ bins. Let $P(M^{(n)})$ denote the resulting empirical probability distribution. We then calculate the Shannon entropy as $h^{(n)} = H(P(M^{(n)})) = -\sum_{i=1}^{K} p_i^{(n)} \log p_i^{(n)}$, where $p_i^{(n)}$ is the probability associated with bin, $i$, and $K$ is the number of bins with nonzero probability. Higher entropy indicates a more dispersed distribution of weekly movement values, whereas lower entropy indicates that movement values are more concentrated in a smaller number of bins.

\section{Results}\label{sec:results}

We now  present our main findings. First, we compare the MLR and RFR models for predicting the time-averaged node degrees for   the network encoding all movements (Sec.~\ref{sec:two_models}) and the network layers associated with different industry types(Sec.~\ref{sec:industry}).
%
%
In Sec.~\ref{sec:Survey}, we compare the performance of all ten predictive models.
%
In Sec.~\ref{sec:Survey_Target}, we turn our attention to the task of predicting weekly node degrees (i.e., as opposed to their time-averaged values).

\subsection{Predicting Time-Averaged Movements Between Census Tracts}\label{sec:two_models}

\begin{figure}[t]
\centering
\includegraphics[width=1\linewidth]{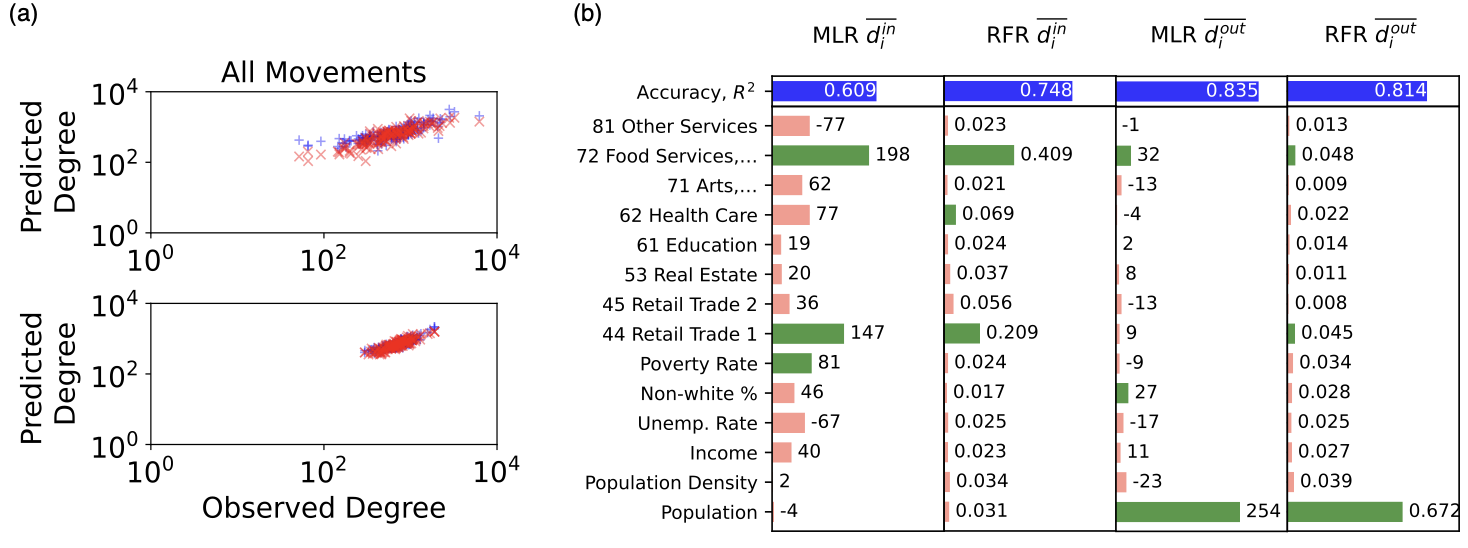}
\includegraphics[width=1\linewidth]{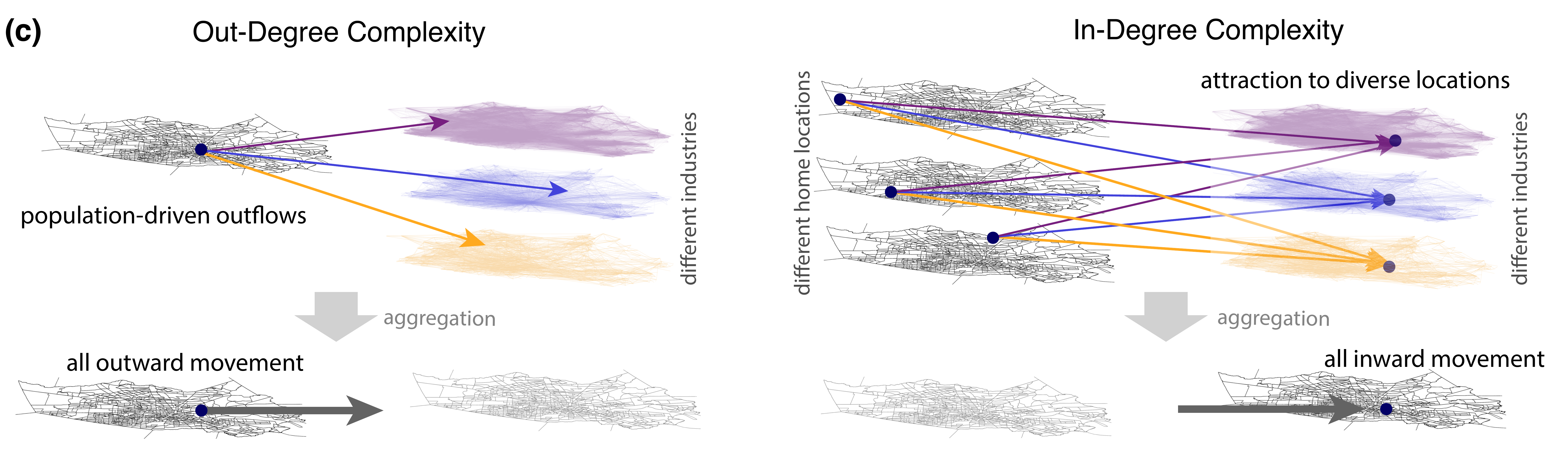}
\caption {
\textbf{Time-averaged node degree prediction using MLR and RFR.}
(a): Predicted vs. observed node degrees for the layer-aggregated network. 
(b):  R-squared values  measure model performance and and the input variables' importance are shown for both models.
(c) Complexity of in- and out- degrees for an industry-stratified mobility network.}
\label{mlr_predicted_vs_observed_regression_mean}
\end{figure}

Here, we compare the MLR and RFR models for predicting the  census tracts' time-averaged in-degrees $\overline{d_i^{\text{in}}}$ and out-degrees $\overline{d_i^{\text{out}}}$ for the network encoding all movements. 
We present our findings in Fig.~\ref{mlr_predicted_vs_observed_regression_mean}. In Fig.~\ref{mlr_predicted_vs_observed_regression_mean}(a),  scatter plots compare observed and predicted values for  $\overline{d_i^{\text{in}}}$ (left) and  $\overline{d_i^{\text{out}}}$ (right).  Different symbols are used for MLR and RFR. Observe most points lie along the diagonal, indicating accurate predictions.

In Fig.~\ref{mlr_predicted_vs_observed_regression_mean}(b), we report the models' R-squared values  quantifying model accuracy (top row) as well as   measures for the input features' importance for each model (lower rows). (Recall Sec.~\ref{sec:predict}.)
Observe the left two columns
correspond to predicting $\overline{d_i^{\text{in}}}$, whereas the third and fourth columns are for  $\overline{d_i^{\text{out}}}$.
We first comment on the R-squared values, which are lower for predicting  $\overline{d_i^{\text{in}}}$ as compared to  $\overline{d_i^{\text{out}}}$, which highlights that it is more difficult to predict movements into census tracts as compared to predicting outward movements.  This was also previously noted in \cite{butler2026multilayer} for the MLR model, and here we find a similar phenomenon for the RFR model. 

Turning our attention to the lower rows  of Fig.~\ref{mlr_predicted_vs_observed_regression_mean}(b), observe for  $\overline{d_i^{\text{out}}}$ predictions (right columns) that the feature opulation is far more important than any other features, for both the MLR and RFR models. That is, outward movements are driven by census tract populations.
A different phenomenon occurs for $\overline{d_i^{\text{in}}}$ predictions, where many input variables are important. The top-two input variables both MLR and RFR models are infrastructure counts for food services and accommodation (72) and retail trade 1 (44). The third most important feature is poverty rate rate (81) for MLR ane and health care (62) for RFR. Thus, no single census tract feature dominates the predictability of inward movements (at least to the extent which population drives outward movement). This phenomenon is illustrated in   Fig.~\ref{mlr_predicted_vs_observed_regression_mean}(c).

\subsection{Predicting Time-Averaged Movements for Separate Industry Sectors}\label{sec:industry}

In Sec.~\ref{sec:two_models}, we studied the prediction of time-averaged in- and  out-degrees for a human mobility network encoding all movements between census tracks. Here, we expand this study to consider the prediction of in- and out-degrees for network layers encoding movements to locations associated with different types of industry sectors. As before, we will compare the MLR and RFR models and study both the prediction accuracy and feature importance.

\begin{figure}[t]
\centering
\includegraphics[width=1\linewidth]{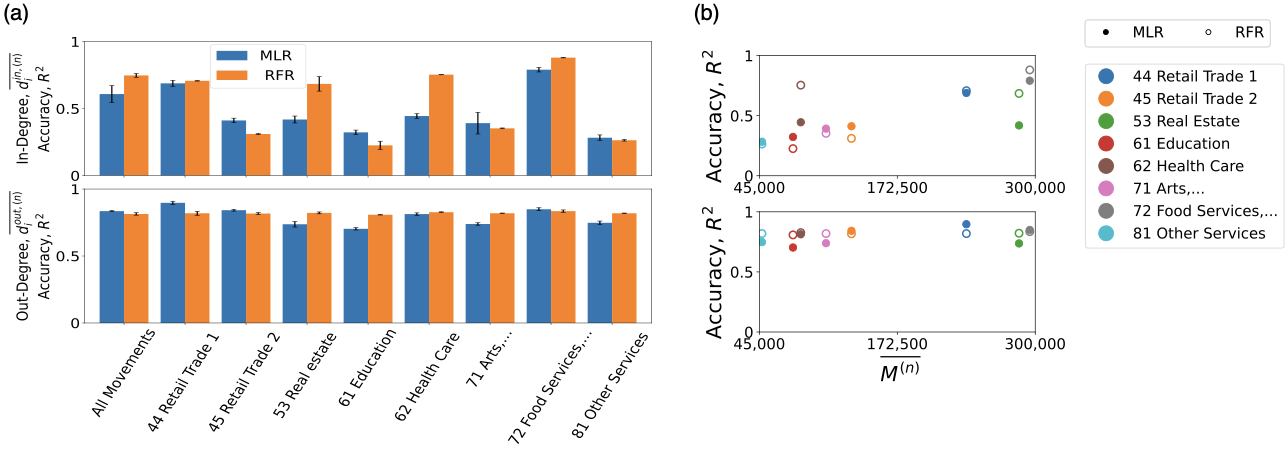}
    \vspace{-.9cm}
    \caption {\textbf{Movement prediction across industry sectors.}    
    (a)~Accuracy across industry sectors for  predictions of time-averaged in- and out-degrees. Error bars illustrate  standard deviations across folds.
    (b)~Prediction accuracy versus time averaged total observed movements for the different industry sectors. }
\label{easier_harder_naics_in_mean_80_20}
\end{figure}

In Fig.~\ref{easier_harder_naics_in_mean_80_20}(a), we report the R-squared values that measure prediction accuracy for node degrees for eight selected industry types(as well those associated with the network encoding all movements categories). The top and bottom panels of Fig.~\ref{easier_harder_naics_in_mean_80_20}(a) correspond to in- and out degrees, respectively. The error bars in Fig.~\ref{easier_harder_naics_in_mean_80_20}(a) represents the standard deviation of the R-squared values across folds; which illustrates the variability of model performance during training. The small error bar suggests that the model's performance fluctuates only slightly across the cross-validation folds. 

First, observe that the R-squared values for out-degree prediction are all higher than those for in-degree prediction, implying that this is a consistent phenomenon characterizing movement predictions across industry types.
Focusing on the in-degree predictions, 
we note that the easiest industry type to predict is food services and accommodation (72), since it has the largest 
R-squared values for both the MLR and RFR models. In contrast, the
hardest industry type to predict is movements to education services (61) for the RFR model and movements to other services (81) for the MLR model.
For the out-degree predictions, we can see that from the bottom panel of Fig.~\ref{easier_harder_naics_in_mean_80_20}(a) that the easiest industry to predict is  retail trade 1 (44) for MLR and  food services and accommodation (72) for RFR. For both models the hardest industry to predict  is education services (61). Finally, it is also interesting to highlight that the MLR and RFR models are generally similar in their predictive capabilities, with the largest difference occurring for the prediction of $\overline{d_i^{\text{in},(n)}}$ for movements to healthcare services (62).

In Fig.~\ref{easier_harder_naics_in_mean_80_20}(b), we examine whether predictability across industry sectors is associated with the amount of movement data for each sector, which is represented by their time averaged total observed movement, $\overline{M^{(n)}}$. The top and bottom panels show the relationship between prediction accuracy and the layers' $\overline{M^{(n)}}$ for time-averaged  degrees $\overline{d_i^{\text{in},(n)}}$ and $\overline{d_i^{\text{out},(n)}}$, respectively. 
Observe for both RFR and MLR models that the R-squared values are strongly correlated with $\overline{M^{(n)}}$ for in-degree prediction. 
For MLR, the Pearson correlation coefficient is $r=0.769$ with a p-value of $p=0.026$. (For RFR $r=0.746$ and $p=0.034$.)  In contrast,  for out-degree prediction the correlations are weaker and not statistically significant (MLR: $r=0.441$, $p=0.275$; RFR: $r=0.540$, $p=0.167$). That is, the predictability of in-degrees for an industry-sector-associated network layer is associated with the number of movements in that layer; in-degree prediction is harder for industry sectors with fewer movements.
  
Next, we investigate which input features are most important for predicting in and out degrees across the network layers associated with different industry types.
In Table~\ref{in_mean_80_20}, we indicate measure of importance for the input features. The top and bottom tables correspond to in- and out-degree prediction, respectively, while the left and right tables correspond to MLR and RFR models, respectively. 
Each row corresponds to a different input variable and each column is a different prediction task, i.e., predicting node degrees for movements associated with different industry industry types. (The R-squared values illustrated in Fig.~\ref{easier_harder_naics_in_mean_80_20}(c) are also provided.)


Observe in Table~\ref{in_mean_80_20}(top) that for a network associated with a particular industry type, it is almost always the case that the most important predictor of in-degrees is the variable that encodes the amount of infrastructure  in each census tract associated with that industry.
This phenomenon is consistent across both MLR and RFR models, except for the MLR-based prediction of $\overline{d_i^{in,(n)}}$ for the network layer encoding movements to locations associated with real estate (53), in which case that input feature is the second-most-important one. Intuitively, this strong pattern is expected, since knowledge about the locations of infrastructure for a particular industry type should be very important for the prediction of movements to those locations. Aside from this pattern, it is also interesting to note that infrastructure information about two industry types---food services and accommodation (72) and retail trade 1 (44)---tend to be important input features even when predicting movements to other industry types. This is consistent with Fig.~\ref{mlr_predicted_vs_observed_regression_mean}, where we found these two variables to be the most important input features for predicting in-degrees for the network encoding all movement categories. Food services and retail stores can increase the attractiveness of an area \cite{oner2017retail} and likely draw human movements from other areas.

\clearpage

\begin{table*}[t]
\caption{{\bf Input feature importance for predicting time-averaged  in- and out-degrees.} The top and bottom tables correspond to in- and out-degree prediction, respectively, while the left and right tables correspond to using the MLR and RFR models, respectively. Within each of the four tables, rows correspond to different input variables, and their importance measures are indicated for the prediction tasks---that is, node-degree prediction for mobility network layers associated with different industry types(as indicated by their NAICS codes). In each column, we highlight the top-three input features with green shading.}
\includegraphics[width=1.0\linewidth]{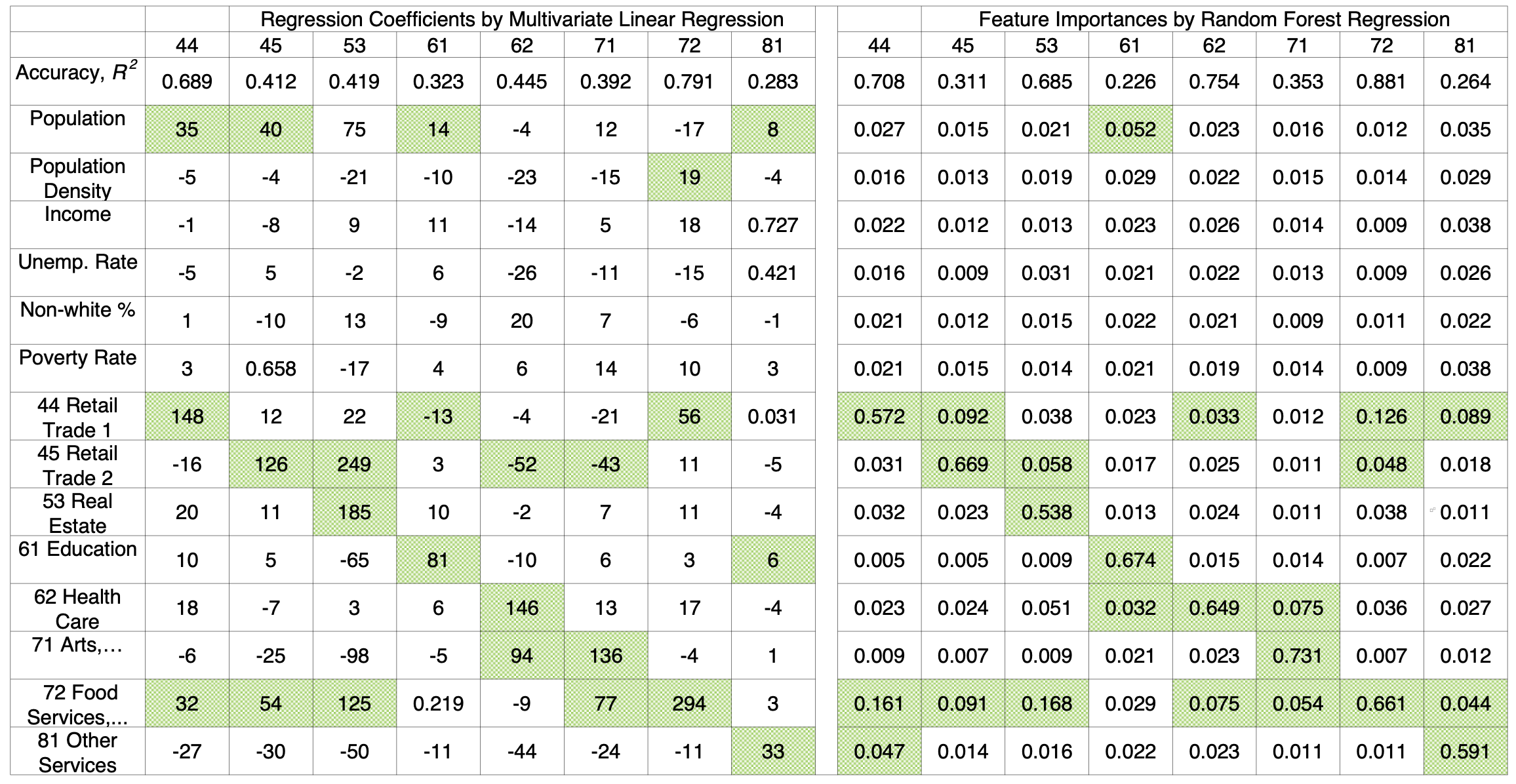}

~

\includegraphics[width=1.0\linewidth]{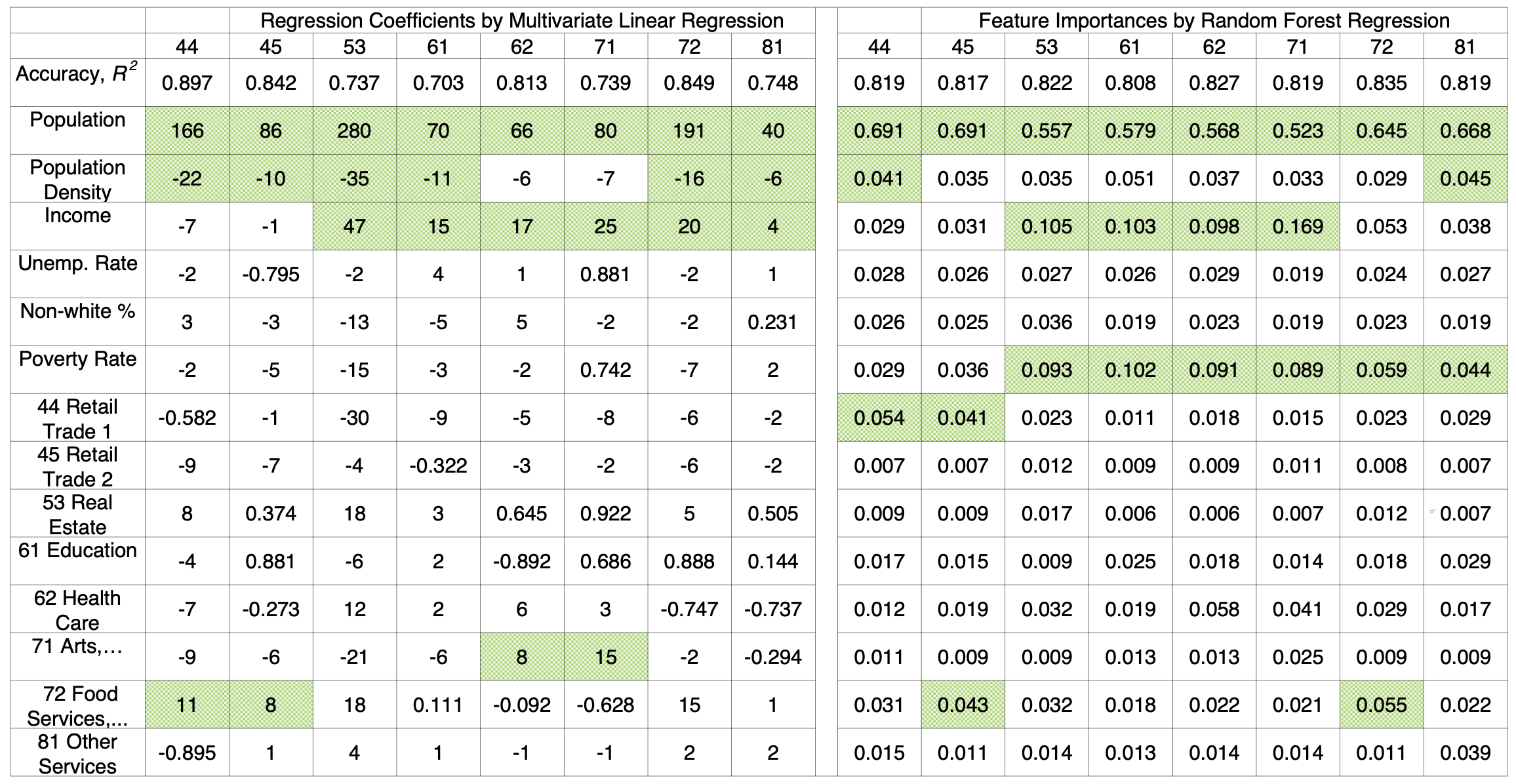}
\label{in_mean_80_20}
\end{table*}

\clearpage

Finally, we study input feature importance for the prediction of nodes' out-degrees across industry typesusing the MLR and RFR methods as shown in Table~\ref{in_mean_80_20}(bottom). Observe for both  models that population is the most important input variable for predicting out-degrees across all  industry types. This is consistent with out findings in Sec.~\ref{sec:two_models} for the network encoding all movements. Income is also consistently important. Interestingly, population density arises as a third important input feature across industry types for MLR, whereas for RFR, it is poverty rate.

\begin{figure*}[b]
\centering
\includegraphics[width=.9\linewidth]{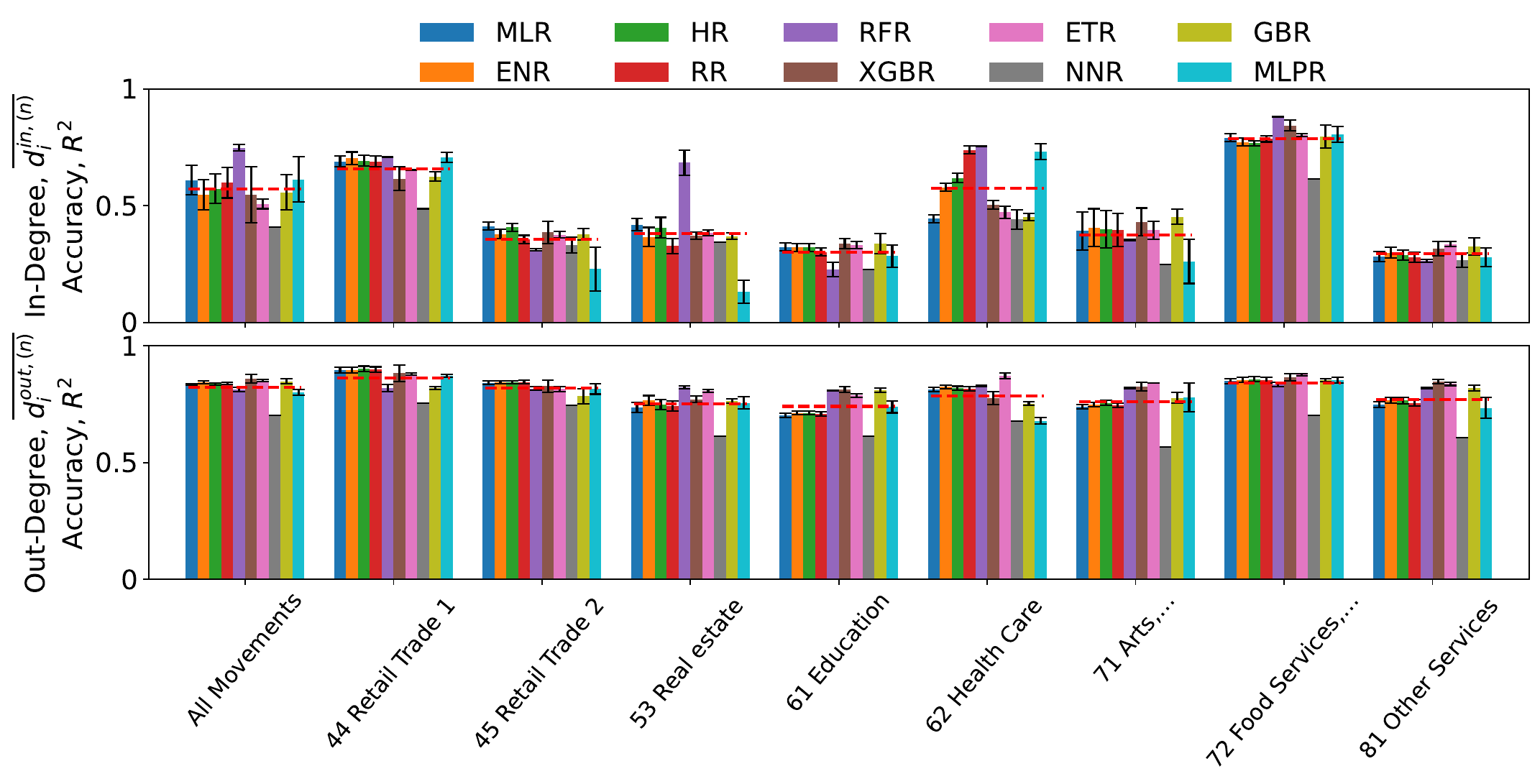}
\vspace{-.2cm}
\caption{\textbf{Comparison of statistical and machine learning models for predicting time-averaged node degrees.}
R-squared values measure model performance for four linear models (the first four) and six nonlinear ones (those remaining). The top and bottom panels indicate $R^2$ for predicting nodes' in- and out-degrees, respectively, and values are given for the network encoding all movements as well as the eight network layers encoding movements to distinct industry types. Black vertical lines illustrate the standard deviation of the R-squared values across folds, and the red horizontal dashed lines indicate for each industry type, the mean value across the ten models. } 
    \label{compare_in_out_ml}
\end{figure*}

\subsection{Comparison Across Ten Predictive Models}
\label{sec:Survey}

In Sections~\ref{sec:two_models} and \ref{sec:industry}, we identified which input variables are most important, and we compared the performance of node degree prediction using two  models: MLR (a classical linear model) and RFR (a nonlinear machine learning model). 
To extend this analysis, we now study eight additional predictive models, and we group them all into two categories.
The linear group includes MLR, ElasticNet Regression (ENR), Ridge Regression (RR), and Huber Regression (HR). The nonlinear group consists of RFR, XGBoost Regression (XGBR), Extra Trees Regression (ETR), Nearest-Neighbor Regression (NNR), Gradient Boosting Regression (GBR), and Multilayer Perceptron Regression (MLPR).
See Sec.~\ref{sec:predict} and Table.~\ref{ML_table} for further discussion about our implementation of these models.



In Fig.~\ref{compare_in_out_ml}, we expand the results shown earlier in Fig.~\ref{easier_harder_naics_in_mean_80_20}(a) by depicting the R-squared values for predicting $\overline{d_i^{in,(n)}}$ and $\overline{d_i^{out,(n)}}$ across industry sectors and across the ten models. 
Similar to before, we observe in-degree prediction to be consistently more difficult than out-degree prediction, regardless of industry type of predictive model used. Also note that the error bars are again small, indicating low variability in model performance across the cross-validation folds. 

Focusing on the top panel of Fig.~\ref{compare_in_out_ml} ($\overline{d_i^{in,(n)}}$ prediction), we highlight for each industry type (and the aggregated network) that the different predictive models can significantly vary in their performance.
When considering the average performance across all cases, RFR emerges as the best-performing model overall, while NNR exhibits the weakest performance. (The mean and standard deviation across models and industry sectors are summarized in the Supplementary Materials.)



Additionally, we study how the $R^2$ values change from one industry to another, which we study by considering for each industry the average performance across all ten models which we plot as red dotted lines in Fig.~\ref{compare_in_out_ml}. 
These aggregated statistics provide a general indication of which sectors are easier or more difficult to predict. Extending our earlier findings, we now find that food services and accommodation (72) ($R^2=0.786$) is the easiest sector to predict in-degrees across all ten predictive models, whereas other services (81) ($R^2=0.294$) is the most challenging industry sector to predict in-degrees. It is worth noting that model-specific behavior may differ. For example, education (61) is the most difficult sector for the RFR model, and the sector of other services (81) is the one that is second-most difficult to predict.

Turning our attention to $\overline{d_i^{out,(n)}}$ prediction, we can observe in the  bottom panel of Fig.~\ref{compare_in_out_ml} that for all industry sectors, there is less variation across the different predictive models' performance (i.e., as compared to the variation for $\overline{d_i^{in,(n)}}$ prediction). Again we study the average performance of each model across industry sectors to find that ETR performs best, and NNR again yields the weakest results. Again, we study the average model performance for each industry sector to find that   retail trade 1 (44) ($R^2=0.862$) is the easiest to predict out-degrees, while the hardest one is education (61) ($R^2=0.741$). (See Supplementary Materials.)
Moreover, model-specific differences also remain evident, since, for example, 
food services and accommodation (72) is the easiest sector to predict out-degrees for the RFR model. 
%
In summary,  while we find that different industry sectors are generally easier/harder to predict, particularly with respect to the harder task of in-degree prediction, the relative performance of different models can vary across prediction tasks and there is no clear model ``winner'' (although RFR performs the best overall).


\subsection{Predicting the Weekly Varying Movements}\label{sec:Survey_Target}

We have so far focused on developing models to predict  $\overline{d_i^{in,(n)}}$ and $\overline{d_i^{out,(n)}}$---that is, the time-averaged in- and out-degrees for each census tract---and we now turn our attention to predictions for the week-to-week node degrees ${d_i^{in,(n)}}(t)$ and ${d_i^{out,(n)}}(t)$. Similar to our study in Sec.~\ref{sec:Survey}, we again consider the performance of ten machine learning models organized into linear and nonlinear groups. 
To allow the models to potentially learn seasonal (i.e., annually varying) affects, we train the models using the same fourteen input features as well as two time-encoding variables that track the 52-week cycles during the two years of data from ($2018$ and $2019$). That is, we define
two additional periodic features given by 
$\sin{\frac{2\pi w}{52}}$
and $\cos{\frac{2\pi w}{52}}$; where $w\in\{1,\dots,52\}$ denotes the week of the year.

\begin{figure*}[!ht]
\centering
\includegraphics[width=.9\linewidth]{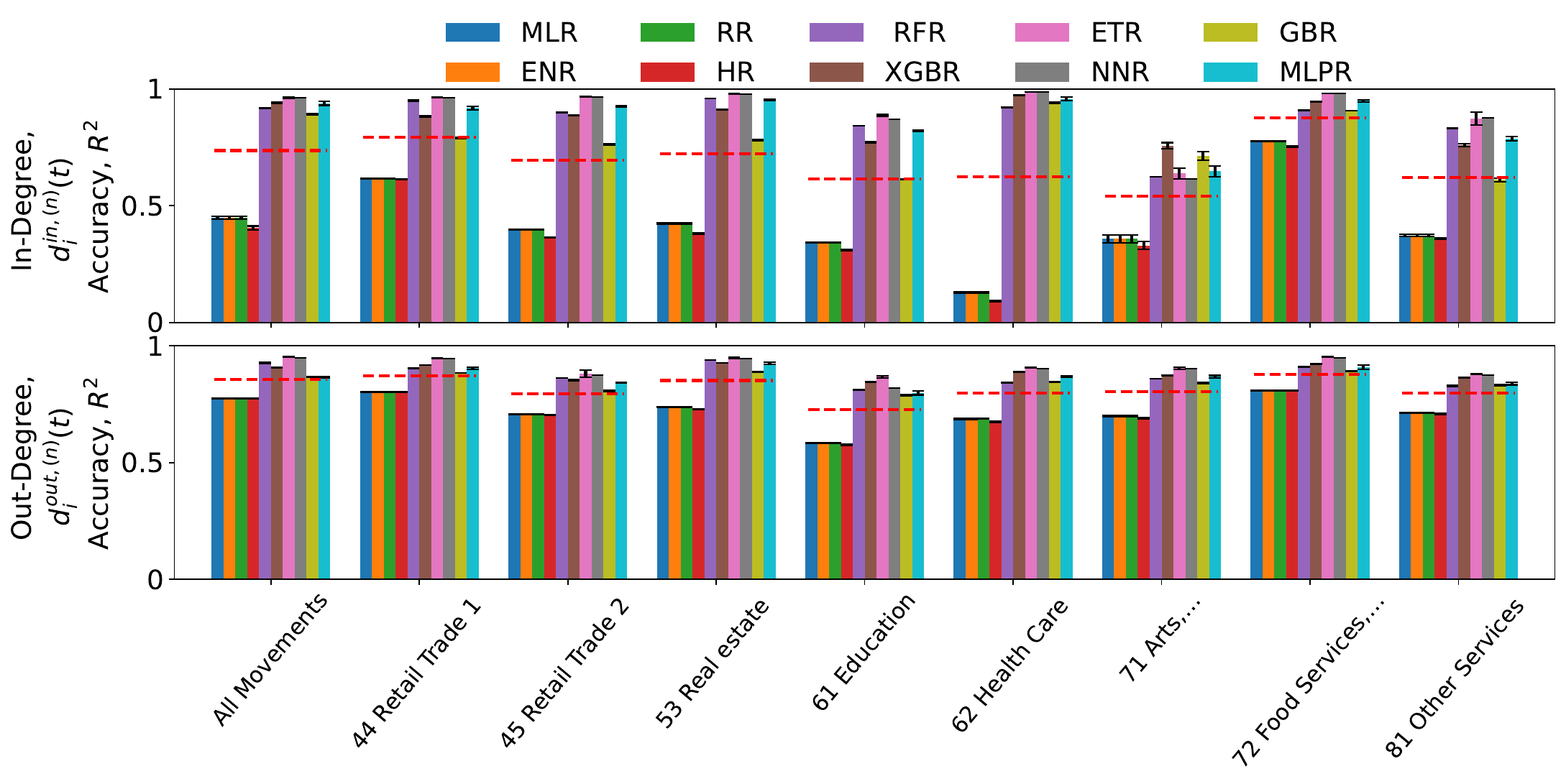}
    \caption {\textbf{Performance of predictive models for weekly in- and out-degree prediction.} R-squared values measure model performance for four linear models (the first four) and six nonlinear ones (those remaining). The top and bottom panels indicate $R^2$ for predicting in- and out-degrees, respectively, and values are given for the network encoding all movements as well as the eight network layers encoding movements to distinct industry types. 
    }
     \label{cyclic_in_out_ml}
\end{figure*}

In Fig.~\ref{cyclic_in_out_ml}, we show results that are analogous to those in Fig.~\ref{compare_in_out_ml}, except we now compare the performance of all ten models for the task of predicting weekly node degrees. That is, we plot the models' R-squared values for predicting
$d_i^{{\text{in},(n)}}(t)$ (top panel) and $d_i^{{\text{out},(n)}}(t)$ (bottom panel).  Red dotted lines indicate mean performance across models for each industry type, and error bars indicate the standard deviation across folds. As before, these measures are given for the network encoding all movements and the network layers associated with different industry sectors. 

Consistent with our earlier findings,
the prediction of $d_i^{{\text{in},(n)}}(t)$ is again found to be more challenging than that of  $d_i^{{\text{out},(n)}}(t)$.
However, if one compares Fig.~\ref{cyclic_in_out_ml} to Fig.~\ref{compare_in_out_ml}, we can observe a significant increase in the predictability of $d_i^{{\text{in},(n)}}(t)$  for the nonlinear machine learning models---their performance was approximately $R^2\approx0.548$ when predicting time-averaged in-degrees, but this increased to $R^2\approx0.915$ when predicting weekly in-degrees. The R-squared values are also much larger now for out-degree predictions. 
%
%
Importantly, the increased R-squared values for the nonlinear models is consistent across all all industry types as well as the network encoding all movement categories. That is, the weekly varying network layers contain more variability (specifically temporal variations), which the nonlinear models can effectively learn.


In the top panel of Fig.~\ref{cyclic_in_out_ml}, one can observe that the predictive performance for $d_i^{{\text{in},(n)}} (t)$ varies substantially across models, indicating sensitivity to model choice. In contrast, for $d_i^{\text{out},(n)}(t)$ prediction (Fig.~\ref{cyclic_in_out_ml}, bottom panel), the performance is more consistent across different machine learning models. For both cases, however, the linear models exhibit nearly identical results.
%
Using the same averaging procedure described in Section.~\ref{sec:Survey}, for $d_i^{\text{in},(n)}(t)$ prediction we find that, among all of the models, ETR achieves the best overall performance whereas HR performs the worst. Furthermore, by averaging across models for each industry type, food services and accommodation (72) ($R^2=0.875)$ is consistently the easiest to predict, while arts and entertainment (71) ($R^2=0.540)$ is the most challenging. (See Supplementary Materials.)

Similarly, one can observe for   $d_i^{{\text{out},(n)}}$ prediction in the bottom panel of Fig.~\ref{cyclic_in_out_ml} that repeatedly ETR achieves the best performance, whereas HR again performs the worst. Based on the averaged results across models, food services and accommodation (72) ($R^2=0.877)$ remains the easiest sector to predict, while  education (61) ($R^2=0.726)$ is the most challenging. (See Supplementary Materials.)

\begin{figure*}[!ht]
\centering
\includegraphics[width=1\linewidth]{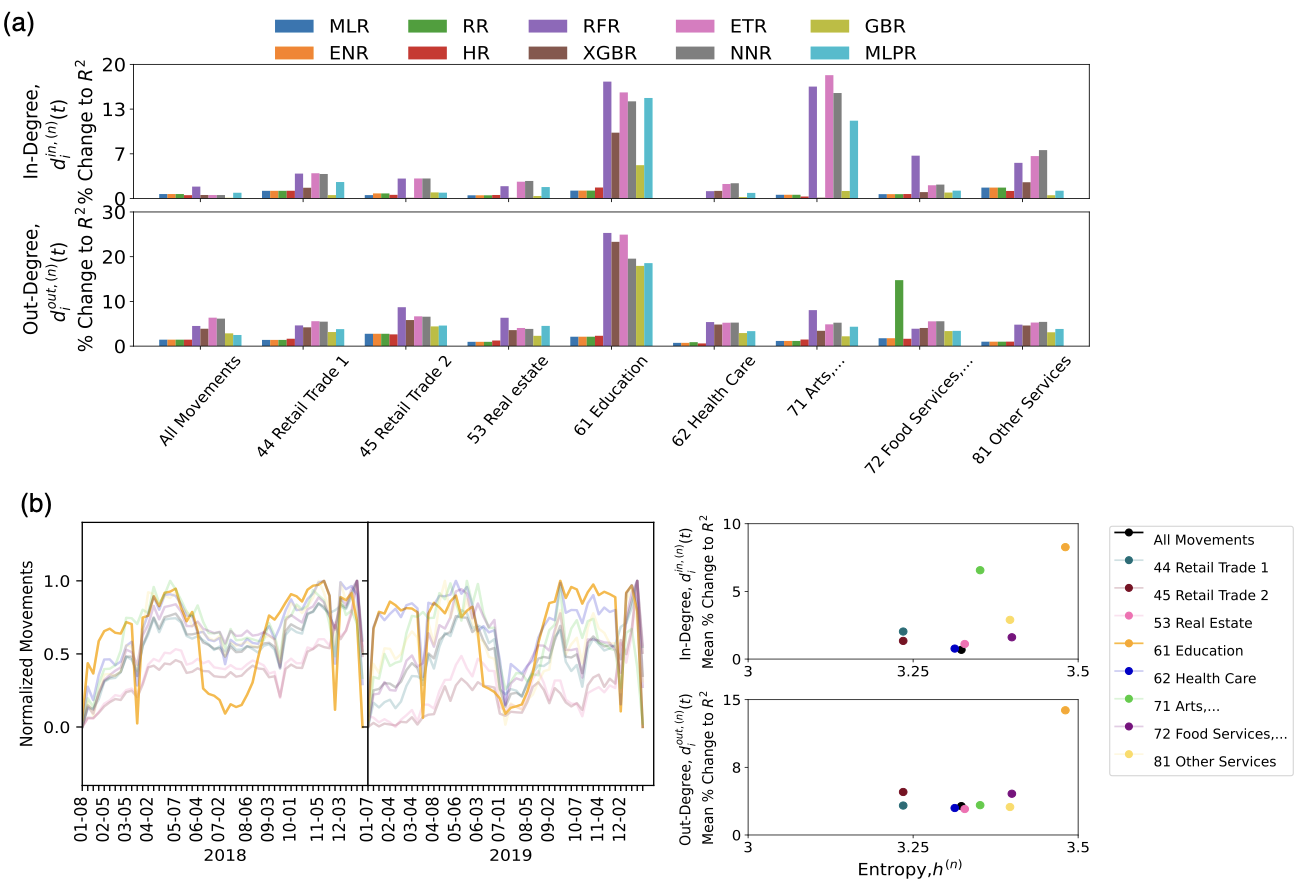}
\caption {\textbf{Importance of time-encoding variables across industry sectors.} 
(a) The experiment giving rise to Fig.~\ref{cyclic_in_out_ml} was repeated without using the time-encoding variables as part of the input data, and here we plot the percentage decrease for the models' R-square values across industry types. All drop in performance. 
(b) Left panel: We plot the amount of movement associated with each industry type each week during the two-year study. To help  visibility, we normalize each curve to have a maximum of 1 and minimum of 0. Right panel: For each industry sector, we compare the percentage change versus the Shannon entropy measuring the complexity for that sectors'  total movements (see Sec.~\ref{sec:entropy}).
} 
    \label{percentage_in_out_ml}
\end{figure*}

One significant difference between our predictions for time-averaged node degrees and weekly node degrees is that the latter includes two additional input variables: the two periodic variable encoding for week of the year. With this in mind, we now study the importance of these variables for enabling the increased prediction accuracy shown in Fig.~\ref{cyclic_in_out_ml}. Specifically, we conducted an experiment  considering the same prediction tasks as before, except that we now omit the time-encoding variables from the input data.

We present our findings in Fig.~\ref{percentage_in_out_ml}. In Fig.~\ref{percentage_in_out_ml}(a), we plot the percentage decrease to the R-squared values shown in Fig.~\ref{cyclic_in_out_ml} that occur when the time-encoding variables are omitted. The top panel indicates the percentage decrease to R-squared value for $d_i^{{\text{in},(n)}}(t)$ prediction, whereas the bottom panel presents the corresponding decrease for $d_i^{{\text{out},(n)}}(t)$ prediction. In both cases, we observe that the linear models are very slightly effected, which highlights that they do not effectively take advantage and utilize the temporal-encoding variables.  In contrast, the nonlinear models' performances are more strongly impacted, which highlights their ability to take advantage of temporal information to improve movement predictions.

Also observe  in Fig.~\ref{percentage_in_out_ml}(a) that node degree predictions for the network layer encoding movements to education services (61) are the most impacted by the omission of the time-encoding input variables. (The R-squared values decreased up to 18\% for weekly in-degree predictions and  25\% for out-degree predictions.) We hypothesize that one cause is that    movements to schools have a more complicated dependence on time versus other industry sectors.

To support this prediction, in the panel of Fig.~\ref{percentage_in_out_ml}(b) we plot the weekly amount of observed movement for each industry sector, $M^{(n)}(t)$,  during the 104-week time period from $2018-2019$. (To improve visibility, each curve is normalized to have a minimum of 0 and maximum of 1.) Observe that the curve associated with educational services appears to be the most complex, which is expected due to the school closures that occur during  the annual breaks for summer, winter, and spring break. To quantify this complexity, we use the Shannon entropy approach described in Sec.~\ref{sec:entropy} to obtain for each network layer $n$ an associated complexity measure $h^{(n)}$. In the right panel of Fig.~\ref{percentage_in_out_ml}(b),
we plot the mean percentage change from Fig.~\ref{percentage_in_out_ml}(a) (i.e., averaging across ML models) versus the entropy measure $h^{(n)}$ for each industry type. As expected,  $h^{(n)}$ is largest for the education sector (61), which exhibits the largest change for both in- and out-degree prediction. In fact, across all industry sectors we observe a positive correlation between $h^{(n)}$ and the percentage change (which as you recall, measures the importance of the time-encoding variables). For in-degree change, the Pearson correlation coefficient is $r=0.632$ with p-value $p=0.068$ (while  $r=0.635$ and $p=0.066$ for out-degree prediction). Overall,  this trend indicates that sectors with greater temporal variability tend to show a larger reduction in predictive performance when the time-encoding variables are omitted.

\section{Discussion}\label{sec:discuss}

In summary, we investigated the predictability of human movement patterns in Harris County, Texas,  by constructing a time-dependent multilayer network in which node represents census tracts and edges capture movements between them. Each network layer encodes a different movement category according to industry sector (e.g., visit to educational service, health sector, retail trade). This approach for network construction follows that which we developed in previous work \cite{butler2026multilayer}, although we emphasize that that prior paper  focused on studying how a natural disaster impacted human mobility, and we currently study movements during a stable period from 2018-2019. Nevertheless, aligning our study with the existing paper has provided an opportunity to compare results to prior MLR-based predictions in the same region.

Our approach to studying movement prediction focused on the nodes' in- and out-degrees that encode movements in and out of census tracts each week (see Sec.~\ref{sec:Survey_Target}), although we first predicted their time-averaged values (see Secs.~\ref{sec:two_models}, \ref{sec:industry}, \ref{sec:Survey}). By predicting inward and outward movements for multilayer networks in which different layers encode movements to locations associated with different industry types, our work complements the existing machine learning literature on movement prediction \citep{cuttone2018understanding,kim2018method,toch2019analyzing,wang2019urban,luca2021survey}. This multilayer network approach also extends  the long-standing literature on  spatial interaction modeling \citep{wilson1971family,fotheringham1989spatial}.


We trained predictive models for node degrees using demographic, socioeconomic, and infrastructure features, and we developed both linear and nonlinear predictive models. Code used in this paper is available on GitHub \citep{GitHub}. Consistent across all experiments, we find that inward-movement prediction is more challenging than outward-movement prediction. This asymmetry suggests that outgoing mobility patterns were more easily captured by the available input features (i.e., predominantly population), whereas incoming mobility patterns were inherently more heterogeneous and difficult to model (recall Fig.~\ref{mlr_predicted_vs_observed_regression_mean}). 
%
This perspective was supported by our  feature importance analysis in Sec.~\ref{sec:two_models}, where we observed out-degree predict to be driven by census tract population, whereas several input features were important for in-degree prediction.


Our industry-sector-based analysis in Sec.~\ref{sec:industry} also revealed that the most accurate predictions were for the network layer encoding movements to food services and accommodations, whereas predictions were hardest for the layers associated with other services  and education.
Interestingly, as illustrated in Fig.~\ref{easier_harder_naics_in_mean_80_20}, the difference in in-degree predictability across industry sectors was found to be strongly correlated with the total amount of movement for each sector.

In Sec.~\ref{sec:Survey_Target}, we focused on predicting weekly movements (as opposed to time-averaged ones) and  incorporated two  temporal features to encode the week of the year.
While our findings for weekly node degrees were similar to those for time-averaged node degrees, we did observe a significant increase in the R-squared values measuring accuracy for in-degree prediction.
The most striking difference, however, was that the accuracy of the nonlinear machine learning models significantly increased for   in-degree predictions, taking advantage of the new time-encoding variables---especially for industry sectors with strong seasonal dynamics. 

Our study of node-degree predictability highlights how machine learning can be used to understand the structural organization of time-varying, multilayer networks encoding human mobility. It also has broad potential applications.
For example, if movements to industry sectors can be predicted, organizations can better anticipate changes in demand and allocate their resources more efficiently. Similarly, by identifying locations and sectors where demand remains steady over time, transportation companies can use mobility predictions to guide infrastructure projects and adjust bus or train schedules. 
In public health and emergency response, knowing which locations and sectors show steady or changing movement patterns can help identify high-risk areas and inform decisions about when and where to direct response efforts.

In future work, it would be interesting to extend our study to investigate other regions, including both highly populated and less populated areas, to better understand the effects of population density on predictability.  It would also be interesting to explore the impact of spatial scale on movement predictability by studying networks with different spatial resolutions.
Additionally, a more detailed investigation of sector-specific predictability could provide further insights into demand fluctuations across industries and broaden our understanding of why certain sectors are inherently easier or more difficult to model. It would also be beneficial to extend our study by considering other machine learning tasks and models. Examples might include predicting edge weights rather than node degrees, using graph neural networks, and formulating node-degree prediction as a time-series prediction task.



\section*{Conflict of Interest Statement}
The authors declared no conflict of interest.

\section*{Funding}
This work was supported by the U.S. National Science Foundation under Grant No. DMS-2401276 and the Jay Kemmerer WORTH Institute for seed funding. Any opinions, findings, and conclusions or recommendations expressed in this material are those of the authors and do not necessarily reflect the views of the National Science Foundation.
\section*{Acknowledgments}
The authors  thank SafeGraph for providing anonymized mobile phone location data. We also thank Chen Xu and Jian Gong for their many discussions on mobility network modeling.

\bibliographystyle{Frontiers-Harvard}
\bibliography{bibliography}




\onecolumn 
\section*{Supplementary Material}

\section*{Extended Results for the Study in Sec.~3.2}\label{observed_vs_predicted_mean}

\begin{figure*}[!ht]
\centering
\includegraphics[width=1\linewidth]{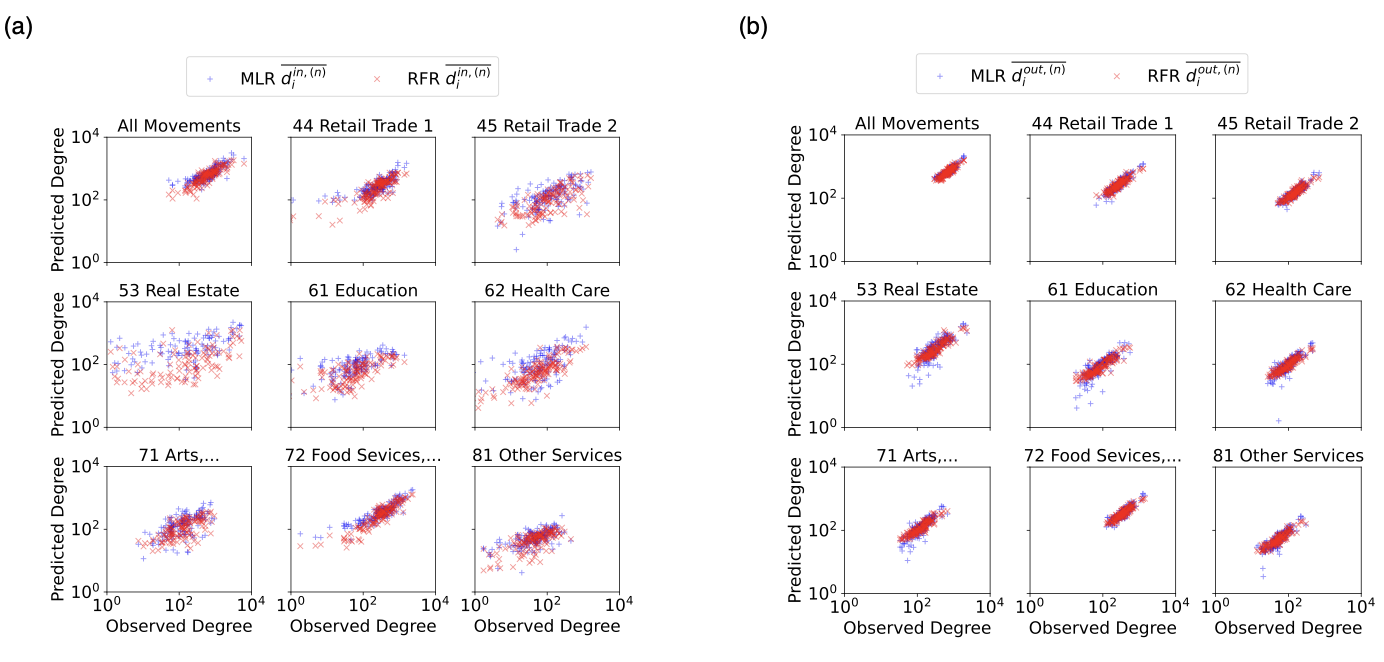}
\caption{(a)-(b) \textbf{Predicted vs. observed values for for time- averaged in- and out-degrees for network layers encoding movements to different industry sectors.}   }
\label{observed_predicted_mean}
\end{figure*}

In Sec. 3.2 of the main text, we investigated the prediction of time-averaged in- and out-degrees using MLR and RFR models for network layers representing movements to locations associated with different industry types.
In Fig.~\ref{observed_predicted_mean}, we compare the observed and predicted values of $\overline{d_i^{\text{in},(n)}}$   and $\overline{d_i^{\text{out},(n)}}$   for different industry types using MLR and RFR. Observe the points mostly fall along a diagonal line. Also observethat across industry types that the out-degree predictions are accurate than those for in-degrees.

\clearpage

\section*{Extended Results for the Study in  Sec.~3.3}\label{Appendix_Table_mean}

In Sec. 3.3, we compared the performance of node degree prediction using ten machine learning models. The following tables provide the numerical results corresponding to the model performance for time-averaged in- and out-degree prediction (i.e., the numerical values associated with Fig.~6 in the main text). The tables also include the  average and standard deviation across models and across  industry sectors. 

\begin{table*}[!h]
\caption{\textbf{Numerical values plotted in Fig.~6 for time-averaged in- and out-degree prediction.} The top and bottom tables associated with in- and out-degree prediction respectively along with the mean and standard deviation (STD) across models (right two columns) and across industry sectors (bottom two rows).}
\label{Fig_6_Appendix_In_Out}
\includegraphics[width=1\linewidth]{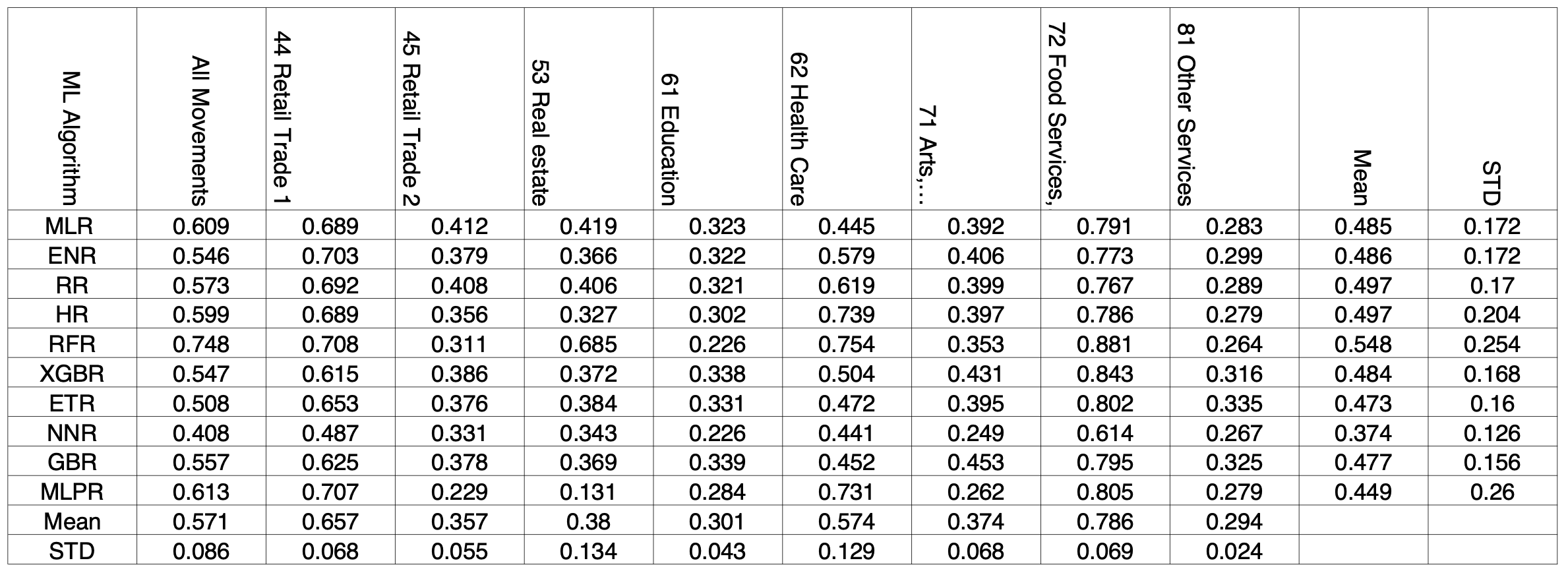}
~
\includegraphics[width=1\linewidth]{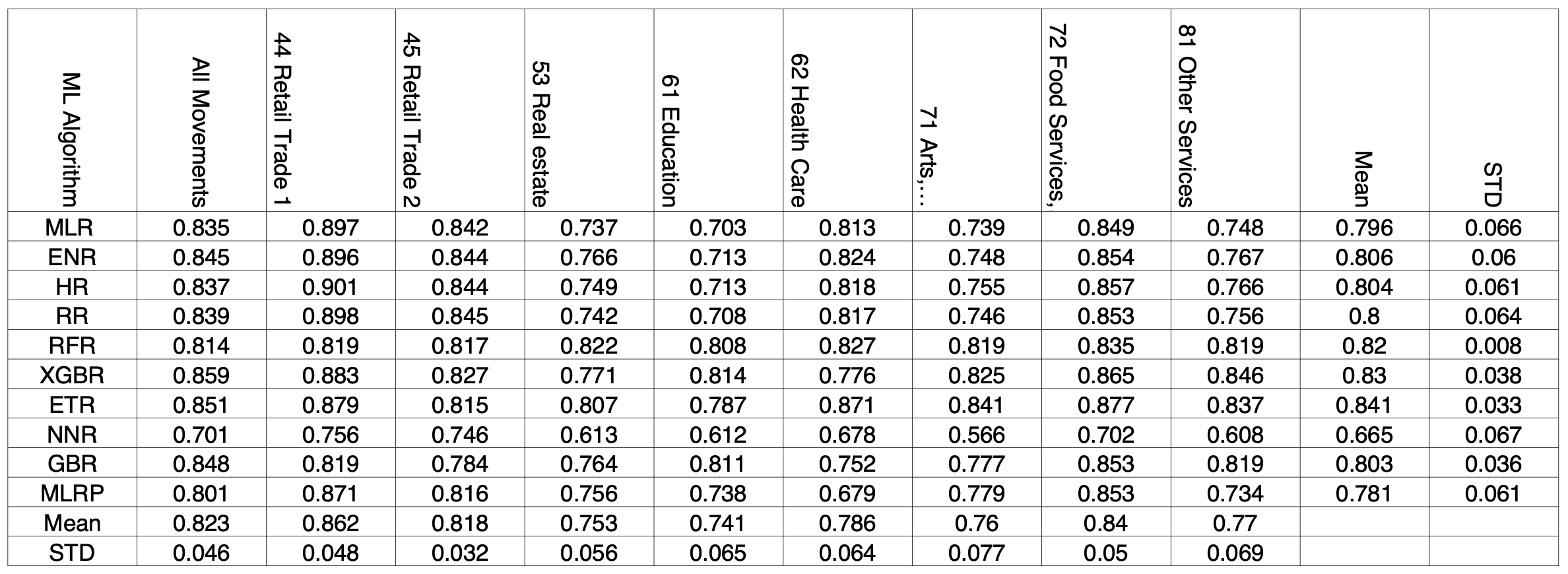}
\end{table*}

\clearpage

\section*{Extended Results for the Study in Sec. 3.4} \label{observed_vs_predicted_weekly}
\begin{figure*}[h]
\centering
\includegraphics[width=1\linewidth]{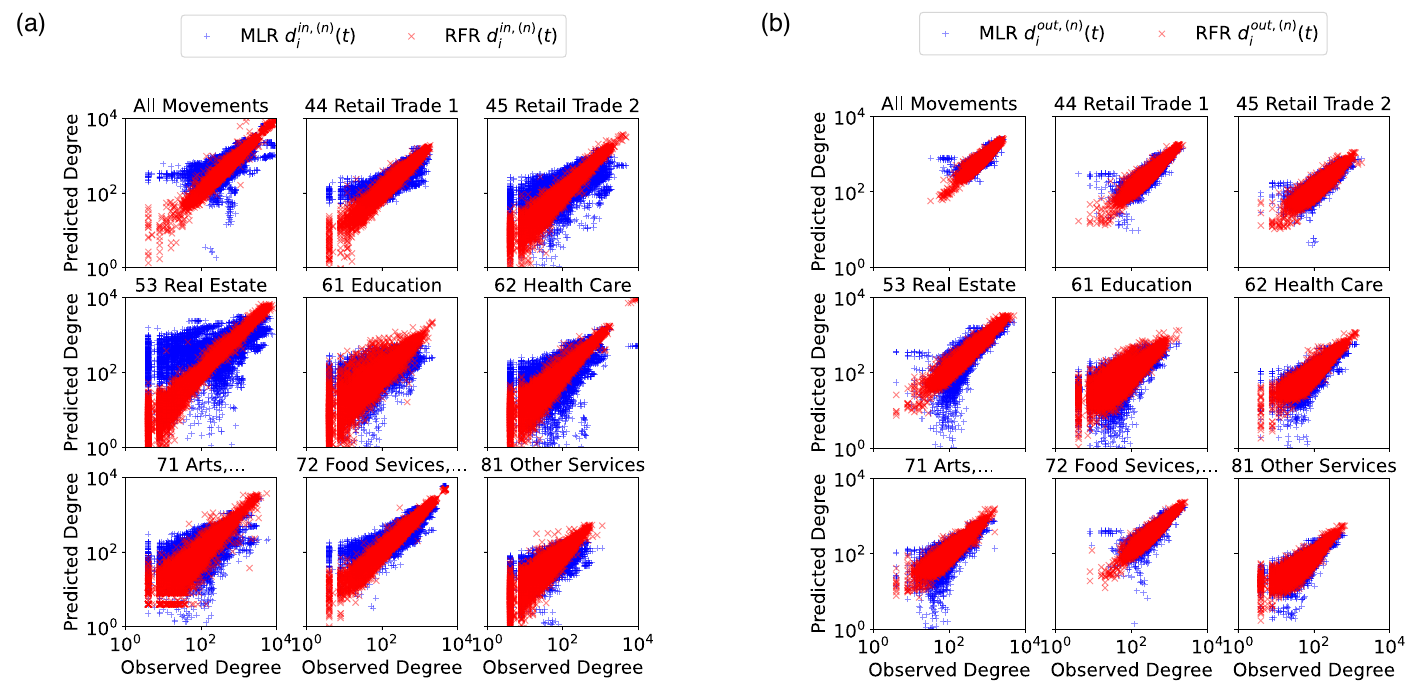}
\caption{(a)-(b) \textbf{Predicted vs. observed values for  census tracts' weekly node degrees.} 
}
\label{observed_predicted_weekly}
\end{figure*}

In Sec. 3.4, we predicted weekly in- and out- degrees including time encoding variables.  
In Fig.~\ref{observed_predicted_weekly}, we compare the observed and predicted in- and out-degrees ${d_i^{in,(n)}}(t)$ and  ${d_i^{out,(n)}}(t)$, respectively. Results are shown for both the MLR and RFR models, and observe the RFR model is much more accurate (i.e., the points more strongly align with the diagonal). Comparing this figure to Fig.~\ref{observed_predicted_mean}, we can observe that the weekly degrees are much more heterogeneous than the time-averaged degrees predictions. Intuitively, this is expected, since averaging across time removes the data variability in that domain.

Next we provide tables below that contain  the numerical values corresponding to Fig.~7 in the paper.

\begin{table*}[!ht]
\caption{\textbf{Numerical values corresponding to  Fig.~7 for weekly node degree prediction.}, The top and bottom tables report weekly in- and out-degree respectively with the mean and standard deviation (STD) across models (right two columns) and across industry types. (bottom two rows).} 
\label{Fig_7_Appendix_In_Out}
\includegraphics[width=1\linewidth]{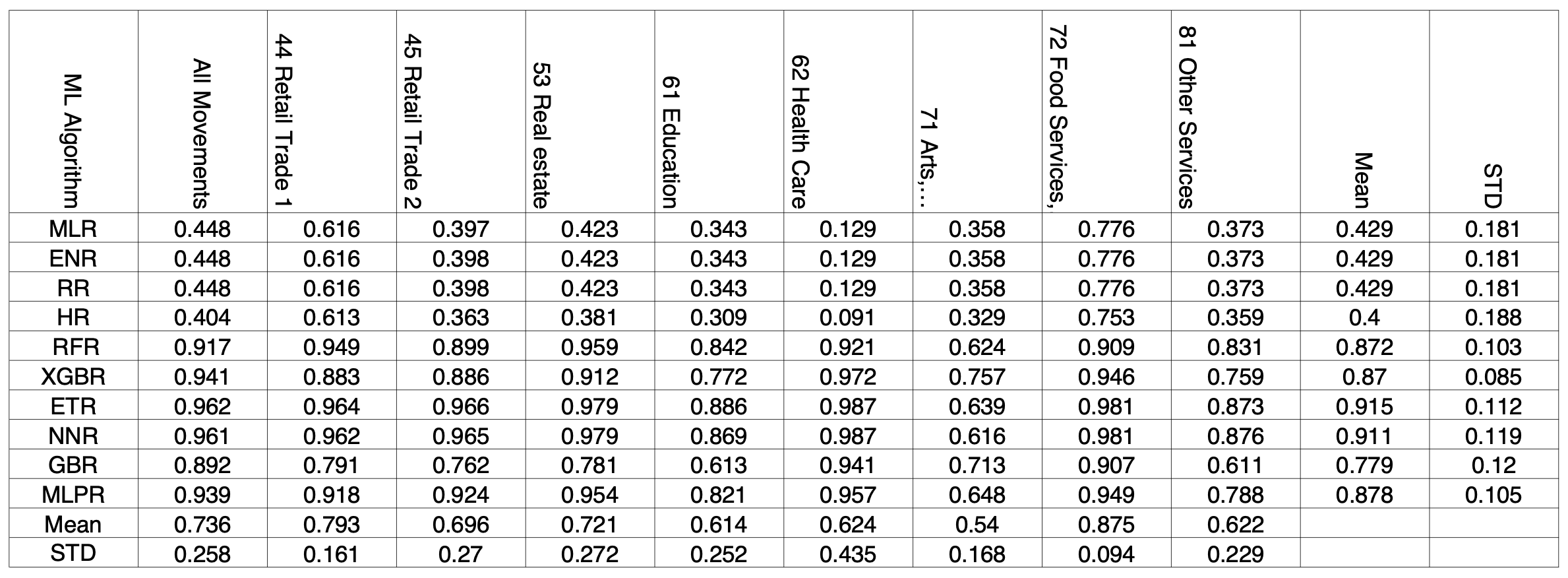}
~
\includegraphics[width=1\linewidth]{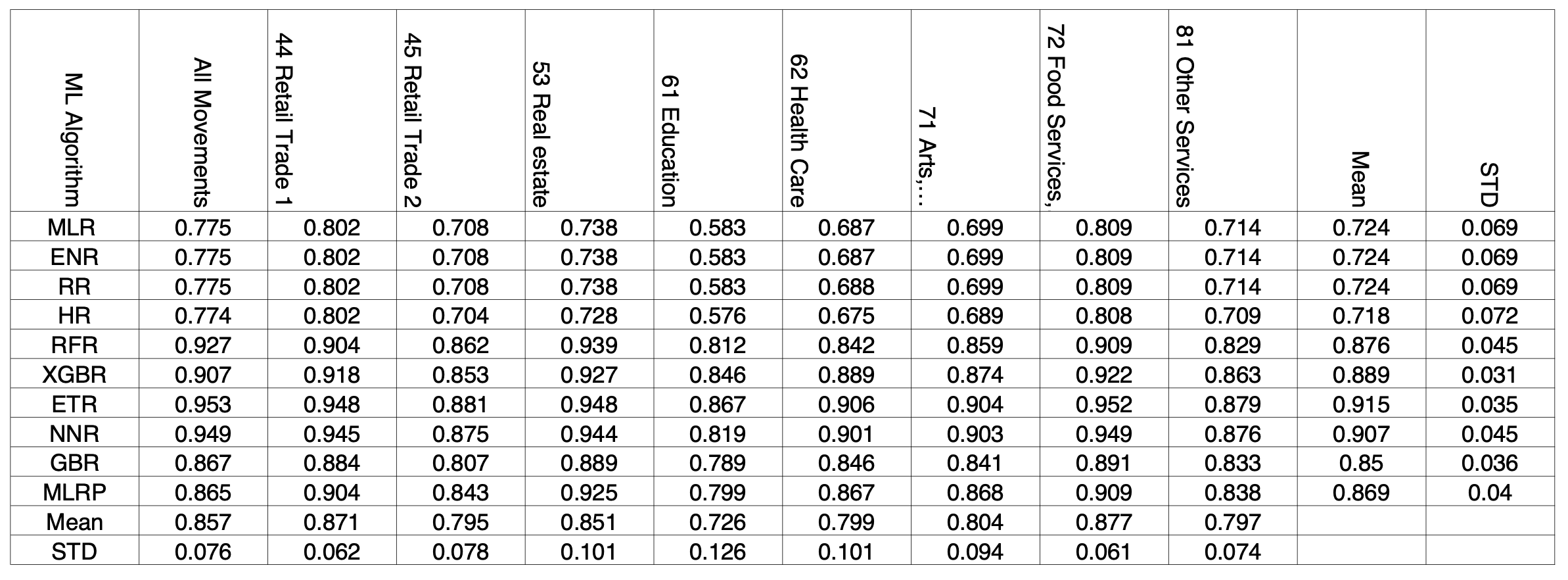}

\end{table*}

\end{document}